\documentclass[11pt,a4paper]{article}
\usepackage{amsmath,amssymb,graphicx}

\def\mydate{February 16, 2006}

\newcommand{\beeq}{\begin{equation}}
\newcommand{\eneq}{\end{equation}}
\newcommand{\beqn}{\begin{eqnarray}}
\newcommand{\eeqn}{\end{eqnarray}}

\def\mybig{\displaystyle \strut }
\def\mbig{\displaystyle }
\def\dd{\partial}
\def\ep{\epsilon}
\def\la{\raise.16ex\hbox{$\langle$}\lower.16ex\hbox{}  }
\def\ra{\, \raise.16ex\hbox{$\rangle$}\lower.16ex\hbox{} }
\def\go{\rightarrow}

\def\onehalf{ \hbox{${1\over 2}$} }

\def\tr{{\rm tr \,}}
\def\eff{{\rm eff}}

\def\myfrac#1#2{{\mybig #1\over \mybig #2}}
\def\mfrac#1#2{{\mbig #1\over \mbig #2}}

\begin{document} 

{\small \noindent \mydate    \hfill OU-HET 557/2006}

\baselineskip 14pt

\begin{center}

\centerline{\LARGE \bf
Physics Consequences of}

\vskip .2cm 

\centerline{\LARGE \bf
Extra-Dimensional Gauge-Higgs Unification\footnote{Proceeding for the  
workshop  {\it ``Fundamental Problems and Applications of Quantum Field Theory''},
YITP, Kyoto, December 19 - 23, 2005.}}

\vspace{2mm}

\centerline{Yutaka Hosotani, Osaka University}

\centerline{\small \tt e-mail:hosotani@phys.sci.osaka-u.ac.jp}
\end{center}

\baselineskip=14pt

In the dynamical gauge-Higgs unification of electroweak interactions, 
the 4D Higgs field is identified with the low energy mode of 
the extra-dimensional component of the gauge potentials.\cite{HM, HNSS, HNT2}
In the Randall-Sundrum warped spacetime, in particular, 
the Higgs boson mass is predicted in the 
range 120 GeV -- 290 GeV, provided that the spacetime structure
is determined at the Planck scale.
Further, couplings of quarks and
leptons to gauge bosons and their Kaluza-Klein (KK) excited states
are determined by the masses of quarks and leptons. 
All quarks and leptons other than top quarks have very small
couplings to the KK excited states of gauge bosons.
The universality of weak interactions is slightly broken
by  magnitudes of $10^{-8}$, $10^{-6}$ and $10^{-2}$  
for $\mu$-$e$, $\tau$-$e$ and 
$t$-$e$, respectively.  

The metric of the Randall-Sundrum warped spacetime is given by
\beeq
 ds^2 = G_{MN}dx^M dx^N
 = e^{-2\sigma(y)}\eta_{\mu\nu}dx^\mu dx^\nu+dy^2, \label{RSmetric}
\eneq
where $\eta_{\mu\nu}={\rm diag~}(-1,1,1,1)$,  $\sigma(y)=\sigma(y + 2\pi R)$, 
and $\sigma(y)\equiv k |y|$ for $|y| \leq \pi R$  
with $k$ being the inverse AdS curvature radius.   
The phenomenology of the Higgs field and charged current interactions of
quarks and leptons is well described by the $SU(3)$ model, though the
enlargement of the gauge group is necessary to correctly describe
neutral current interactions.
In the $SU(3)$ model the relevant part of the action is 
\beeq
I = \int d^5 x \sqrt{-G} ~
\bigg[  -\frac{1}{2} \tr  F^{MN}F_{MN}
+i\bar{\psi} {e_A}^N \Gamma^A D_N \psi -iM \ep \bar{\psi}\psi   \bigg] ~, 
 \label{cL}
\eneq
where $G \equiv \det(G_{MN}) $ and
$\Gamma^A$ is a 5D $\gamma$-matrix.
$D_N \psi = (\dd_N -\frac{1}{8} \omega_{NAB} [\Gamma^A, \Gamma^B] 
-ig_5 A_N ) \psi$ for a triplet $\psi$ and $\ep (y) = \sigma'(y)/k$. 
$M$ is a kink mass parameter.  The boundary conditions are given by
$A_\mu(x,-y) = P_0 A_\mu(x,y)P_0^{-1}$, 
$A_\mu(x,\pi R+y) = P_\pi A_\mu(x,\pi R-y) P_\pi^{-1}$, etc.
We take $P_0 = P_\pi = {\rm diag~}(-1,-1,+1)$. The residual symmetry 
is $SU(2) \times U(1)$.  

The Wilson line phase (the non-Abelian Aharonov-Bohm phase) is given by
\beeq
 \theta_W  = 
  g_5 \int_0^{\pi R} dy  \,   A_y^7(y)   ~~.
\label{Wilson1}
\eneq
The 4D Higgs field corresponds to fluctuations of $\theta_W$.  The value of
$\theta_W$ in the vacuum is determined by the location of the absolute minimum of
the effective potential $V_\eff(\theta_W)$. When $\theta_W \not= 0$ ($mod~\pi$),
the dynamical electroweak symmetry breaking takes place and the symmetry 
breaks down to $U(1)_{EM}$.

\bigskip
\leftline{\large \bf The Kaluza-Klein mass scale, 
the Higgs mass \& quartic coupling constant}

\vskip 5pt

The Kaluza-Klein mass scale is given by 
\beeq
m_{KK} = \myfrac{\pi k}{e^{\pi kR} - 1} = 
\begin{cases}
1/R &\hbox{for } k \go 0,\cr
\noalign{\kern 3pt}
\pi k e^{-\pi kR} 
   &\hbox{for } kR> 2.
\end{cases}
\eneq
The mass of the $W$-boson is given by 
\beeq
 m_W \sim \frac{m_{\rm KK}}{\pi}
 \Big( \frac{2}{\pi kR} \Big)^{1/2} 
 \Big| \sin \frac{\theta_W}{2} \Big|
\eneq
for $e^{\pi kR} \gg 1$.  With $m_W$ and  $k = O(M_{pl})$ given,  one finds,
for generic values of $0.1 \pi < |\theta_W| < 0.9 \pi$,  that 
$kR = 12 \pm 0.1$. 

The effective potential $V_\eff(\theta_W)$ is estimated to be
\beeq
 V_{\rm eff}(\theta_W) = \frac{3}{128\pi^6}m_{\rm KK}^4 f(\theta_W), 
\eneq
where $f(\theta_W)$ is a dimensionless periodic function of $\theta_W$ 
with a period $2\pi$. 
The explicit form of $f(\theta_W)$ depends on the matter content of the model, 
but its typical size is of order one 
in the minimal model or its minimal extension. 

By expanding  $V_{\rm eff}(\theta_W)$ around its global minimum, the 
mass and self-coupling of the 4D Higgs field are determined. 
The results are summarized in Table 1.  There appears an enhancement
factor $\onehalf \pi kR \sim 18.8$ in the Randall-Sundrum spacetime.

\def\myvt{$\vphantom{\myfrac{1}{2}}$}
\def\mvt{$\vphantom{\mfrac{1}{2}}$}

\begin{table}[h]
\begin{center}
\begin{tabular}{|c||c|c|} 
\hline
{\bf ~} & {\bf flat spacetime} & {\bf Randall-Sundrum spacetime} \\
\hline 
~$m_{KK}$~ & $\myfrac{\pi m_W}{\onehalf |\theta_W|}$ 
   & $\myfrac{\pi m_W}{|\sin \onehalf \theta_W|} 
     \Big( \myfrac{\pi kR}{2} \Big)^{1/2}$  \\ 
\mvt & $320 \sim 800$ GeV & $1.5 \sim 3.5$ TeV \\ \hline
$m_H$ & ~$\Big(\myfrac{3}{32\pi} f^{(2)}(\theta_W) \alpha_W \Big)^{1/2}
 \myfrac{m_W}{\onehalf |\theta_W|}$~ 
   & ~$\Big(\myfrac{3}{32\pi} f^{(2)}(\theta_W) \alpha_W \Big)^{1/2}
 \myfrac{m_W}{|\sin \onehalf \theta_W|} \myfrac{\pi kR}{2}$~  \\ 
\mvt & $6 \sim 15$ GeV & $125 \sim 290$ GeV \\ \hline
$\lambda$ & $\myfrac{1}{16} \alpha_W^2 f^{(4)}(\theta_W)$ 
   & $\myfrac{1}{16} \alpha_W^2 f^{(4)}(\theta_W)
    \Big( \myfrac{\pi kR}{2} \Big)^2$ \\ 
\mvt & $0.0008$ & $0.3$  \\    \hline
\end{tabular}
\end{center}
\vskip -.3cm 
\caption{Kaluza-Klein mass scale, Higgs mass and Higgs quartic 
coupling constant.  Numerical values are for $kR=12$ and
$\theta_W = (0.2 \sim 0.5) \pi$.}
\label{g0_values}
\end{table}

\vskip .2cm

\leftline{\large \bf  Non-universality of weak interactions}

\vskip 5pt

In the dynamical gauge-Higgs unification the magnitude of the 
couplings of quarks and leptons to the $W$-boson depends on flavor
through the value of the kink mass $M$.  
The value of $M/k$ is given by 0.865, 0.715, 0.633 and 0.436 
for $e$, $\mu$, $\tau$ and 
$t$(top), respectively. The universality of the weak interactions
is slightly broken.  The magnitude of the deviation 
from the universality is summarized in Table 2. Improvement 
of experiments is awaited for the confirmation of the prediction.

\vskip -.2cm 
\begin{table}[h]
\begin{center}
$\myfrac{g_{(0)}^f}{g_{(0)}^e} - 1$  ~:~ ~
\begin{tabular}{|c||c|c|c|} 
\noalign{\kern 15pt}
\hline
 \mvt $\theta_W$ & $\mu$(muon) & $\tau$(tau) & $t$(top)  \\ \hline 
\mvt ~$0.2\pi$~ & $- 1.74 \times 10^{-9}$ 
   & $-4.69 \times 10^{-7}$ & $-4.3 \times 10^{-3}$ \\ \hline
\mvt $0.5\pi$ & $- 9.26 \times 10^{-9}$ 
   & $-2.50 \times 10^{-6}$ & $-2.2 \times 10^{-2}$ \\ \hline
\mvt $0.8\pi$ & $- 1.70 \times 10^{-8}$ 
   & $- 4.60 \times 10^{-6}$ & $-4.0 \times 10^{-2}$ \\ \hline
\end{tabular}
\end{center}
\vskip -.3cm
\caption{Non-universality of weak interactions. 
The deviation of $g_{(0)}^f/g_{(0)}^e$ from 1 for 
$f = \mu, \tau, t$ is listed where $g_{(0)}^f$ is the coupling
constant of the fermion $f$ to the $W$-boson. ($kR=12$.)
}
\label{g0_values}
\end{table}

\def\jnl#1#2#3#4{{#1}{\bf #2} (#4) #3}

\def\Zphys{{\em Z.\ Phys.} }
\def\jssc{{\em J.\ Solid State Chem.\ }}
\def\jpsJ{{\em J.\ Phys.\ Soc.\ Japan }}
\def\ptps{{\em Prog.\ Theoret.\ Phys.\ Suppl.\ }}
\def\PTP{{\em Prog.\ Theoret.\ Phys.\  }}

\def\JMP{{\em J. Math.\ Phys.} }
\def\NPB{{\em Nucl.\ Phys.} B}
\def\NP{{\em Nucl.\ Phys.} }
\def\PLB{{\em Phys.\ Lett.} B}
\def\PL{{\em Phys.\ Lett.} }
\def\PRL{\em Phys.\ Rev.\ Lett. }
\def\PRB{{\em Phys.\ Rev.} B}
\def\PRD{{\em Phys.\ Rev.} D}
\def\PRe{{\em Phys.\ Rep.} }
\def\AP{{\em Ann.\ Phys.\ (N.Y.)} }
\def\RMP{{\em Rev.\ Mod.\ Phys.} }
\def\ZPC{{\em Z.\ Phys.} C}
\def\SCI{\em Science}
\def\CMP{\em Comm.\ Math.\ Phys. }
\def\MPLA{{\em Mod.\ Phys.\ Lett.} A}
\def\IJMPA{{\em Int.\ J.\ Mod.\ Phys.} A}
\def\IJMPB{{\em Int.\ J.\ Mod.\ Phys.} B}
\def\EPJC{{\em Eur.\ Phys.\ J.} C}
\def\PR{{\em Phys.\ Rev.} }
\def\JHEP{{\em JHEP} }
\def\cmp{{\em Com.\ Math.\ Phys.}}
\def\JPA{{\em J.\  Phys.} A}
\def\JPG{{\em J.\  Phys.} G}
\def\NJP{{\em New.\ J.\  Phys.} }
\def\CQG{\em Class.\ Quant.\ Grav. }
\def\ATMP{{\em Adv.\ Theoret.\ Math.\ Phys.} }
\def\ibid{{\em ibid.} }

\renewenvironment{thebibliography}[1]
         {\begin{list}{[$\,$\arabic{enumi}$\,$]}  
         {\usecounter{enumi}\setlength{\parsep}{0pt}
          \setlength{\itemsep}{0pt}  \renewcommand{\baselinestretch}{1.2}
          \settowidth
         {\labelwidth}{#1 ~ ~}\sloppy}}{\end{list}}

\vskip .2cm 

\end{document}